\documentclass{article}
\usepackage{spconf,amsmath,graphicx,booktabs,subcaption,float,multirow,tikz,pgfplots}

\title{Efficient Adapters for Giant Speech Models}
\name{Nanxin~Chen, ~~Izhak~Shafran, ~~Yu~Zhang, ~~Chung-Cheng~Chiu, ~~Hagen~Soltau, ~~James~Qin, ~~Yonghui~Wu}
\address{Google DeepMind, USA}
\begin{document}
\maketitle
\begin{abstract}
Large pre-trained speech models are widely used as the de-facto paradigm, especially in scenarios when there is a limited amount of labeled data available.
However, finetuning all parameters from the self-supervised learned model can be computationally expensive, and becomes infeasiable as the size of the model and the number of downstream tasks scales.
In this paper, we propose a novel approach called Two Parallel Adapter (TPA) that is inserted into the conformer-based model pre-trained model instead.
TPA is based on systematic studies of the residual adapter, a popular approach for finetuning a subset of parameters.
We evaluate TPA on various public benchmarks and experiment results demonstrates its superior performance, which is close to the full finetuning on different datasets and speech tasks.
These results show that TPA is an effective and efficient approach for serving large pre-trained speech models.
Ablation studies show that TPA can also be pruned, especially for lower blocks.
\end{abstract}
\begin{keywords}
model adaptation, pretrained model, residual adapter
\end{keywords}
\section{Introduction}
\label{sec:intro}
Recent years researchers observed the prevalent paradigm shift from random initialization to transfer learning using pretrained self-supervised learning models, yield superior performance on many various downstream tasks~\cite{baevski2019vqwav2vec, schneider2019wav2vec, wav2vec2, chung2021w2v, pmlr-v162-chiu22a}.
Self-supervised learning methods learn to extract useful feature representations on large amount of unsupervised data through the objectives such as contrastive learning~\cite{schneider2019wav2vec, wav2vec2}, masked language model loss~\cite{baevski2019vqwav2vec,pmlr-v162-chiu22a}, or both~\cite{chung2021w2v}.
Then the whole network is finetuned with small learning rate to make sure that the learned useful representation could be used on downstream tasks.

This approach introduces two types of potential issues.
First, the whole network needs to be updated during finetuning, which are very costly to train and very costly to store.
In last few years, we observed a large amount of different speech applications based on finetuning those self-supervised models~\cite{krishnan2022self, baevski2022data2vec, pepino2021emotion, jia2021png}.
If one wants to build a service which provides all those different applications, finetuning is not a feasible solution either to store or to serve because all the parameters need to be stored and loaded.
For example, if we want to support a new language with the existing model, it is not realistic to re-train the whole model.
Second, to unlock the potential of existing self-supervised learning, a critical point is to use large network which enables to extract expressive representations.
When this is finetuned on small amount of labelled data, it tends to overfit quickly and without proper configuration, catastrophic forgetting~\cite{french1999catastrophic} happens so the information learned through self-supervised learning is abandoned.

Residual adapter~\cite{houlsby2019parameter, bapna2019simple} provides a neat solution to the two problems mentioned above.
Residual adapter is a very simple network which consists of a residual connection with two fully connected layer including nonlinear activation functions like ReLU~\cite{relu}.
Some previous work~\cite{bapna2019simple} also includes an optional layer norm before the fully connect layer to normalize the input.
When we want to build a model for one task, instead of updating the whole network, we only update the added residual adapters.
Typically the residual adapter is much smaller in comparison to the large frozen encoder.
This solves both the issue of storing and serving, since only the adapter needs to be stored and loaded while the frozen encoder is preserved once.
Because of the small amount of trainable parameters, it is often observed that it overfits less frequently.

In this paper, we provide a systematic study of the various residual adapters following a recent study~\cite{he2022towards} which studies the residual adapters for Transformers~\cite{vaswani2017attention}.
Based on the comprehensive study, we further propose a new approach based on those observations for Conformer~\cite{conformer}, which reports the best performance for various speech tasks~\cite{bigssl}.
We report results on different public benchmarks, including Automatic Speech Recognition (ASR) tasks on multiple English datasets~\cite{speechstew}, multilingual dataset~\cite{conneau2022fleurs}, and Automatic Speech Translation (AST) task~\cite{covost2}.
Our study is based on our large pre-trained speech foundation model, which includes about 2 billion parameters.
We chose a large model for study because it has observed that large model benefits more from self-supervised learning~\cite{wav2vec2, radford2022robust} and it offers impressive representation abilities.
The marginal cost of using a large model with adapters decreases significantly as the number of tasks increases, in contrast to small models which are easy to serve but hard to scale.

The paper is organized as follows.
Section \ref{sec:pre} gives a brief overview of different previous work on residual adapters and related approaches.
Section \ref{sec:resapt} introduces the idea of residual adapter and its variations.
Section \ref{sec:exp} starts with massive experiments on different factors of residual adapters on public benchmark.
Based on those studies, we proposed a new variation, Two Parallel Adapter (TPA), which combines the best configuration from each factor.
We report comprehensive results of the proposed TPA on downstream tasks including ASR and AST.
Section \ref{sec:sum} summaries the paper, discusses the broad impact and future directions.

\section{Previous Works}
\label{sec:pre}
Residual adapters have been broadly studied in prior works.
It was originally proposed for transfer learning in natural language processing~\cite{houlsby2019parameter, bapna2019simple, wang2021k, pfeiffer2021adapterfusion} and then has been applied to other modalities, like speech~\cite{hou2021exploiting, tomanek2021residual}, and cross-modalities~\cite{sung2022vl, pan2022stadapter}.

Our studies are closely related to the previous work~\cite{he2022towards} that conduct experiments on natural language processing tasks and we are applying similar studies on Conformer and speech tasks.
In~\cite{thomas2022efficient}, the authors proposed adding residual adapters to self-supervised speech model and applied it to the problem of speech recognition.
Our work is also related to~\cite{le2021lightweight} where residual adapters are used for speech translation.
They studied two scenarios, including adapting from the pre-trained speech translation backbone and transfer learning with both pre-trained encoder and decoder.
Our work is different in two respects.
First, we propose a new variation of residual adapter, TPA, specially designed for Conformer.
Experiments suggest TPA leads to better performance than original residual adapter.
Second, we utilize a unified framework where pre-trained encoder keeps frozen during training while adapters and randomly initialized decoder are added for different tasks, in contrast to the simple ctc decoder~\cite{thomas2022efficient} or pretrained decoder~\cite{le2021lightweight}.
We verify the effectiveness of our framework on both speech recognition and speech translation benchmarks.

\section{Residual Adapters}
\label{sec:resapt}

Residual adapter is a small neural network component which is inserted to play an role of model adaptation.
Instead of updating all parameters, this small component is finetuned so the functionality from pretraining is largely preserved.
The residual adapter consists of a residual connection, two fully connected layers with a bottleneck dimension, one non-linear activation function, and an optional layer normalization~\cite{ba2016layer} before the first fully connected layer.
The form could be expressed as follows:
\begin{equation}
    \mathrm{adapter}(x) = x + (W_2(f(W_1 g(x) + b_1)) + b_2)
\end{equation}
where $f$ is the nonlinear activation function, g could be layer normalization or identity function.

\begin{figure*}[ht] %
   \begin{subfigure}{0.25\linewidth}
       \includegraphics[width=\linewidth]{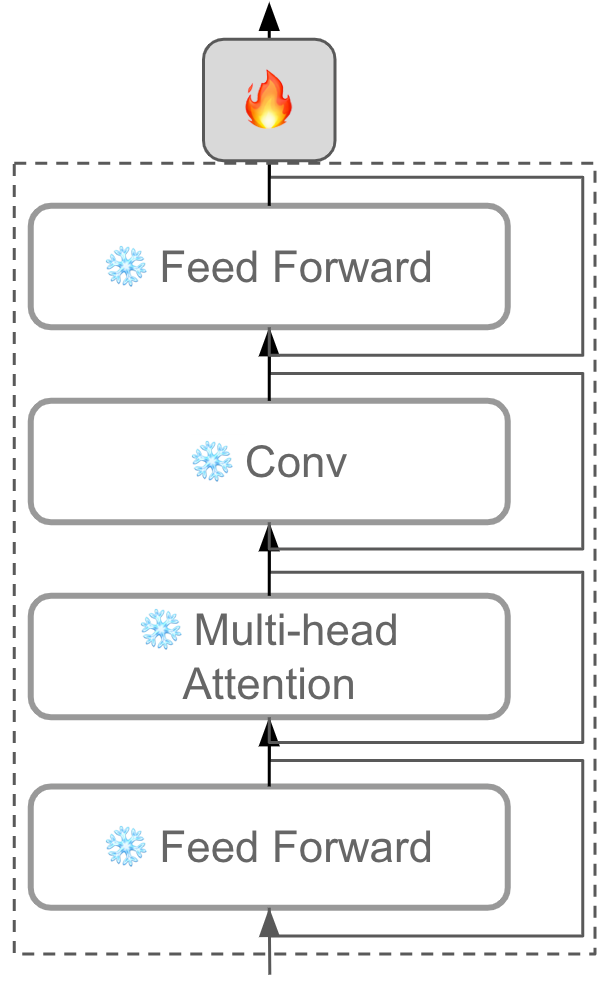}
       \caption{}
       \label{fig:subim1}
   \end{subfigure}
\hfill %
   \begin{subfigure}{0.27\linewidth}
       \includegraphics[width=\linewidth]{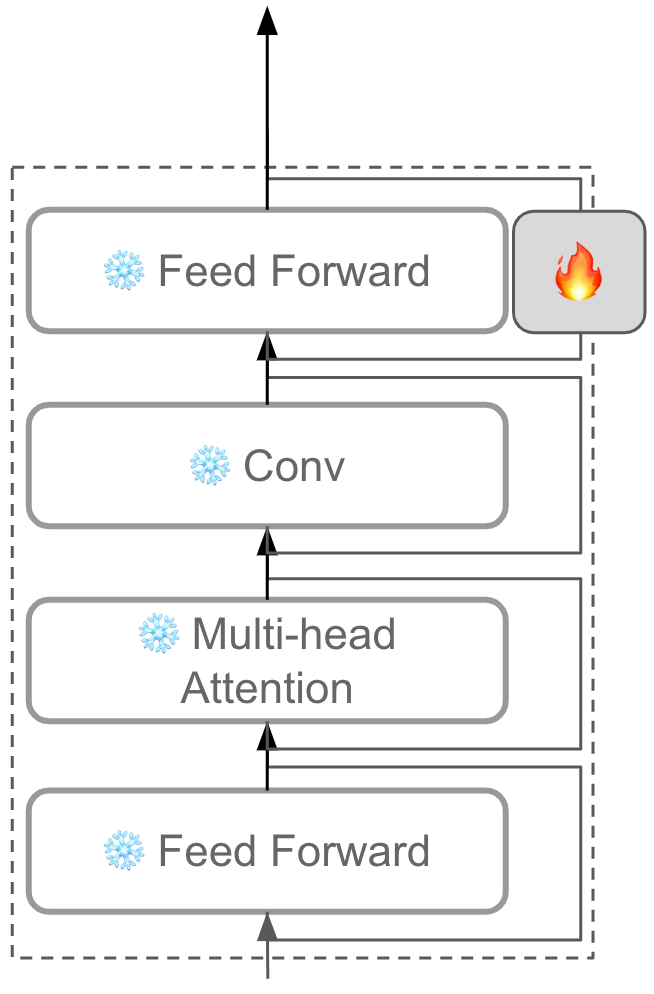}
       \caption{}
       \label{fig:subim2}
   \end{subfigure}
\hfill %
   \begin{subfigure}{0.25\linewidth}
       \includegraphics[width=\linewidth]{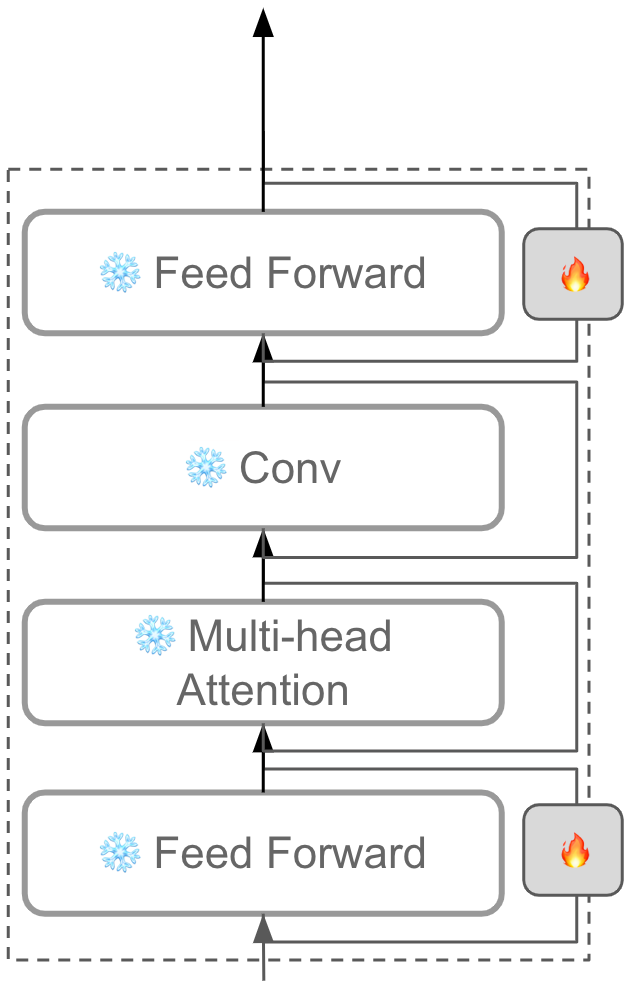}
       \caption{}
       \label{fig:subim3}
   \end{subfigure}

   \caption{Conformer block with different residual adapter variations. Here we demonstrate one conformer block and adapters (represented by the fire icon) are added to all conformer blocks without weight sharing. The left image (\subref{fig:subim1}) represents the serial adapter inserted between conformer blocks. Other two include two parallel adapters studied in this paper. The difference between middle (\subref{fig:subim2}) and right (\subref{fig:subim3}) is to use one wide adapter or two narrow adapter.}
   \label{fig:adapter}
\end{figure*}
In this paper we mainly study two different styles of adapters: {\sl serial} and {\sl parallel}.
We compare results from three different cases, which are shown in Figure~\ref{fig:adapter}.
The first two are the most widely used ones in literature~\cite{bapna2019simple, he2022towards, tomanek2021residual, le2021lightweight}.

For the serial adapter, it is inserted after a component, like an Feed-forward Network (FFN) or conformer block.
For instance, it could be added between conformer blocks as listed in the left part of Figure~\ref{fig:adapter}, and the equation is:
\begin{equation}
    \mathrm{FFN}_{\mathrm{serial}}(x) = \mathrm{adapter}(\mathrm{Conformer}(x))
\end{equation}
since the residual adapter itself includes the residual connection so the output from conformer block is already included.

For the parallel adapter, it is added to the residual connection of the network component, for example, FFN.
Since residual adapter also has a residual connection, the total output becomes the sum of all 3 paths, the input, main branch, and residual adapter:

\begin{equation}
    \mathrm{FFN}_{\mathrm{parallel}}(x) = x + 0.5 * \mathrm{FFN}(x) + (\mathrm{adapter}(x) - x)
\end{equation}
Here we subtract x to avoid counting the input twice.%

The widely used Conformer structure is different in comparison to Transformer and has two FFNs instead of one.
In such case, it is possible to add two adapters for each conformer block which includes two FFNs, as shown in Figure~\ref{fig:subim3}.
We discuss all the possible combinations of those configurations in Experiment part.

\section{Experiments}
\label{sec:exp}

In our experiments, we study the impact of different residual adapter factors and report the best configuration, Two Parallel Adapter (TPA), on various downstream tasks and public benchmarks. 
All our experiments use the same pretrained network, so all results are comparable to each other.
Our pretrained model is based on BEST-RQ~\cite{pmlr-v162-chiu22a} following default masking and quantization parameters.
The model has about 2 billion parameters and includes 32 conformer blocks with hidden dimension 1536.
Layer normalization instead of batch normalization is adopted for training stabilities.
The model is trained on more than 12M hours data covering more than 300 languages.

After pretraining, we add residual adapter to the conformer encoder and finetune it on in-domain supervised data.
For speech recognition tasks, a 2-layer LSTM decoder is randomly initialized with a width of 1280, and it is added to the pre-trained encoder with adapters.
Both the adapter and the decoder are updated with in-domain training data.
For speech translation tasks, we attach a 6-layer, 512-dimension Transformer decoder to the pre-trained encoders.

The residual adapter includes two layers with a bottleneck dimension of 256, and a ReLU non-linear activation function~\cite{relu}, unless otherwise specified.
We initialize $W_2$ with all zeros and $W_1$, $b_1$, $b_2$ with Xavier initialization~\cite{xavier}.

\subsection{Ablation Studies}
We study different factors of residual adapters based on the FLEURS dataset~\cite{conneau2022fleurs}, which is a multilingual dataset that includes 102 languages.
The total number of training data is less than 1000 hours, and the amount of training data varies for different languages.
For example, western European languages have more than 230 hours of training data, while Chinese, Japanese, and Korean (CJK) languages have less than 40 hours of training data.
We believe this dataset represents different cases well and is challenging enough to study different approaches.
We use one adapter for all 102 languages, since it gives better performance in practice.

\subsubsection{Comparison with or without layer norm}
Layer normalization is used in some previous studies~\cite{bapna2019simple,thomas2022efficient} while not used in others~\cite{he2022towards} so we want to study whether this is a crucial factor.
It is introduced in~\cite{bapna2019simple} in order to avoid finetuning existing layer normalization layers within the pre-trained model.
However, the normalization may also be harmful for the adapter learning because it changes the scale of the input.
We used a serial residual adapter, demonstrated in Figure~\ref{fig:subim1}, as baseline system to study whether the layer norm is helpful.

\begin{table}[htbp]
\centering
\caption{Ablation studies on the Layer Normalization. Here the baseline system is shown in Figure~\ref{fig:subim1} and Word Error Rate (WER) is reported grouped by languages with geographical areas: Western
European (WE), Eastern Europe (EE), Central-Asia/MiddleEast/North-Africa (CMN), Sub-Saharan Africa (SSA), South
Asia (SA), South-East Asia (SEA) and Chinese, Japanese, and
Korean (CJK) languages.}
\label{tab:layernorm}
\begin{tabular}{c|ccccccc|c}
 \toprule
 \bfseries System & WE & EE & CMN & SSA & SA & SEA & CJK & Avg \\
 \midrule
 \midrule
 w/ LN & \textbf{15.8} & \textbf{9.9} & \textbf{18.4} & \textbf{33.6} & \textbf{19.4} & \textbf{13.0} & 27.0 & \textbf{19.4} \\
 \midrule
 w/o LN & \textbf{15.8} & \textbf{9.9} & \textbf{18.4} & \textbf{33.6} & 19.7 & 13.2 & \textbf{25.3} & \textbf{19.4} \\
 \bottomrule
\end{tabular}
\end{table}

From Table~\ref{tab:layernorm}, clearly layer normalization doesn't really improve the performance for adaptation overall.
As a result, we don't include any layer normalization layers in the following experiments.

\subsubsection{Comparison between different adapters}

Next, we aim to determine which configuration of residual adapters is best: serial or parallel.
As shown in Figure~\ref{fig:adapter}, for parallel adapters, we study two different possibilities: one where a large adapter is added to one Feed-forward Network (FFN) as Figure~\ref{fig:subim2} and another where two small adapters are attached to both FFNs as~\ref{fig:subim3}.
The intuition is that the Conformer block includes two FFNs and it might be beneficial to adapt both with half-size adapters.
The results are included in Table~\ref{tab:structure}.
Despite some small gains in training speed from better parallelism, the single parallel adapter has very similar performance to the serial adapter.
However, if we consider the architecture of conformer block as shown in Figure~\ref{fig:adapter}, it is possible to attach two smaller adapters to both FFNs.
We call this approach Two Parallel Adapter, or TPA.
TPA reports superior performance compared to the two approaches studied before and obtains the best performance on the Fleurs corpus.
TPA improves performance on all language groups except CJK (Chinese, Japanese, and Korean) which is a small group with large variations.
From the comparison, we can conclude that adapting both FFNs within the conformer block is beneficial and improves performance.

\begin{table}[htbp]
\caption{Ablation studies on different adapter structures listed in Figure~\ref{fig:adapter}. Results are reported in Word Error Rate (WER).}
\label{tab:structure}
\centering
\begin{tabular}{c|ccccccc|c}
 \toprule
 System & WE & EE & CMN & SSA & SA & SEA & CJK & Avg \\
 \midrule
 \midrule
 \begin{tabular}[c]{@{}c@{}}Serial\\(\ref{fig:subim1})\end{tabular} & 15.8 & 9.9 & 18.4 & 33.6 & 19.7 & 13.2 & \textbf{25.3} & 19.4 \\
 \midrule
 \begin{tabular}[c]{@{}c@{}}Parallel\\(\ref{fig:subim2})\end{tabular} & 16.1 & 10.0 & 18.6 & 33.7 & 19.6 & 12.9 & 25.8 & 19.5 \\
 \midrule
 \begin{tabular}[c]{@{}c@{}}Parallel\\(Conv)\end{tabular} & 15.8 & 10.1 & 18.2 & 34.3 & 19.5 & 13.1 & 25.2 & 19.3 \\
 \midrule
 \begin{tabular}[c]{@{}c@{}}TPA\\(\ref{fig:subim3})\end{tabular} & \textbf{15.6} & \textbf{9.8} & \textbf{18.0} & \textbf{33.3} & \textbf{19.0} & \textbf{12.8} & 26.1 & \textbf{19.1} \\
 \bottomrule
\end{tabular}
\end{table}

\subsubsection{Comparison between different sizes}

\begin{figure}[ht]
   \centering
   \includegraphics[width=0.6\linewidth]{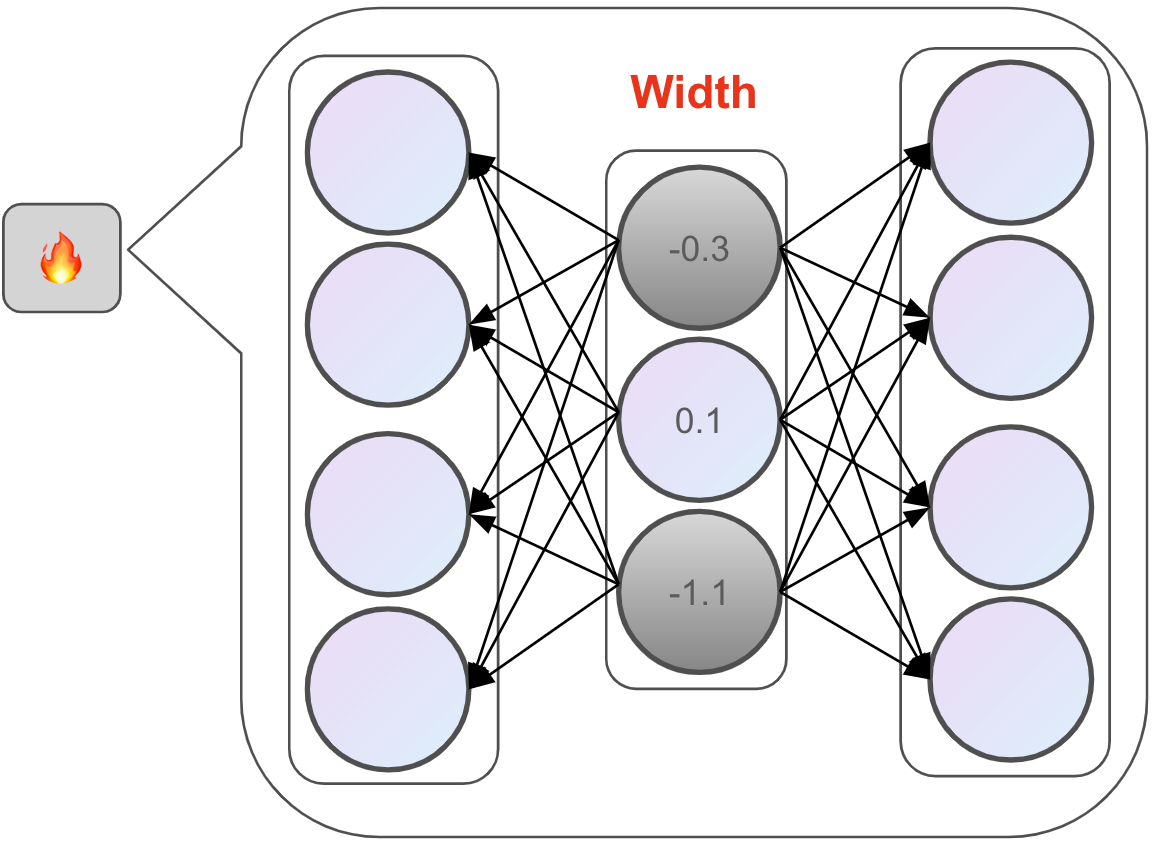}
   \caption{One residual Adapter. Residual connection is not plotted for simplicity. Size of the intermediate layer controls the \textbf{width} for the adapter. Since ReLU activation function is used, only a fraction of those neurons are activated.}
   \label{fig:width}
\end{figure}

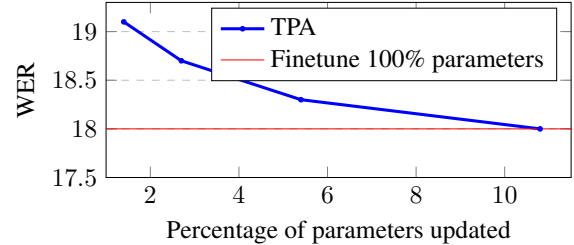
\begin{figure}[t]
  \centering
  \resizebox{0.9\columnwidth}{!}{
  \pgfplotsset{compat=1.3}
\begin{tikzpicture}
\begin{axis}[
    xlabel={Percentage of parameters updated},
    ylabel={WER},
    ymin=17.5, ymax=19.3,
    xmin=1, xmax=11.5,
    legend pos=north east,
    ymajorgrids=true,
    grid style=dashed,
    legend cell align=left,
    width=8cm,height=4cm
]
\addplot[
    color=blue,
    mark=*,
    mark size=0.5,
    very thick,
    ]
    coordinates {
    (1.4, 19.1)(2.7, 18.7)(5.4, 18.3)(10.8, 18.0)
    };
\addplot[mark=none, red] coordinates {(1,18) (100,18)};
    \legend{TPA, Finetune 100\% parameters}
\end{axis}
\end{tikzpicture}}
  \caption{Ablation studies on the width of TPA. The width controls the size of each individual residual adapter and TPA includes two residual adapters per block. When we increase the width, the number of parameters increase and it takes longer time to load.}
    \label{tab:width}
\end{figure}

We further investigate performance enhancement by modifying the number of parameters in the TPA.
The parameter count is controlled by the width of residual adapters which indicates the size of the intermediate layer.
Since for adapters the input and output have the same dimension as the last layer, the size depends totally on the width, as demonstrated in Figure~\ref{fig:width}.
The comparison is provided in Figure~\ref{tab:width}.
Results suggest performance improvement correlated to adapter width. Even on the data-constrained Fleurs dataset, TPA achieved comparable performance to full-network finetuning with only 10\% of the original number of free parameters. Performance degradation was less than 5.6\% with only 1.4\% of the parameters.

\subsection{Analysis}

Based on the results, we want to study if a large width like 1024 is necessary - whether it is possible to shrink the size.
Recall that we use ReLU as the non-linear activation function, as shown in Figure~\ref{fig:width}, it is possible that some intermediate neurons are never activated which means the weighted sum is always non-positive.
In this case, it is safe to remove this neuron and it doesn't change the final result.
Also, it is possible that each language/task only uses a small subset of all neurons as discussed in~\cite{zhang2022moefication}.
We want to study if we could prune the size of the adapter based on the activation statistics.

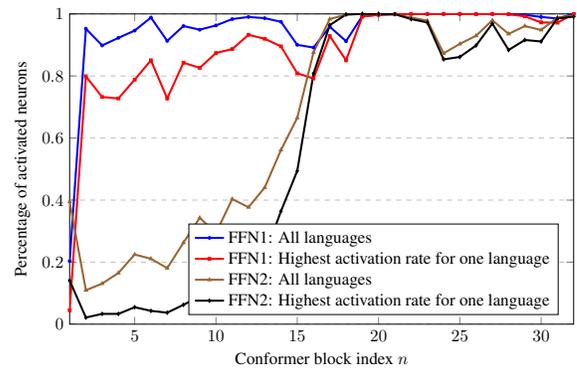
\begin{figure}[t]
  \centering
  \resizebox{0.9\columnwidth}{!}{
  \pgfplotsset{compat=1.3}
\begin{tikzpicture}
\begin{axis}[
    xlabel={Conformer block index $n$},
    ylabel={Percentage of activated neurons},
    ymin=0, ymax=1,
    xmin=1, xmax=32,
    legend pos=south east,
    ymajorgrids=true,
    grid style=dashed,
    legend cell align=left,
    width=12cm,height=8cm
]
\addplot[
    color=blue,
    mark=*,
    mark size=0.5,
    very thick,
    ]
    coordinates {
    (1,0.203125)(2,0.9521484375)(3,0.8984375)(4,0.9228515625)(5,0.9462890625)(6,0.98828125)(7,0.9130859375)(8,0.9609375)(9,0.94921875)(10,0.962890625)(11,0.9833984375)(12,0.990234375)(13,0.986328125)(14,0.974609375)(15,0.900390625)(16,0.8916015625)(17,0.958984375)(18,0.912109375)(19,0.9931640625)(20,0.9970703125)(21,0.9990234375)(22,1.0)(23,1.0)(24,1.0)(25,1.0)(26,1.0)(27,1.0)(28,0.9990234375)(29,0.9970703125)(30,0.990234375)(31,0.9853515625)(32,1.0)
    };
\addplot[
    color=red,
    mark=square*,
    mark size=0.5,
    very thick,
    ]
    coordinates {
    (1,0.044921875)(2,0.798828125)(3,0.732421875)(4,0.7275390625)(5,0.7880859375)(6,0.8505859375)(7,0.7275390625)(8,0.8427734375)(9,0.826171875)(10,0.8740234375)(11,0.88671875)(12,0.9326171875)(13,0.919921875)(14,0.8955078125)(15,0.80859375)(16,0.7919921875)(17,0.9287109375)(18,0.8505859375)(19,0.9912109375)(20,0.9970703125)(21,0.9990234375)(22,1.0)(23,1.0)(24,1.0)(25,1.0)(26,0.9990234375)(27,1.0)(28,0.9990234375)(29,0.9921875)(30,0.97265625)(31,0.9716796875)(32,1.0)
    };
\addplot[
    color=brown!80!black,
    mark=triangle*,
    mark size=0.5,
    very thick,
    ]
    coordinates {
    (1,0.39453125)(2,0.109375)(3,0.130859375)(4,0.1650390625)(5,0.224609375)(6,0.2109375)(7,0.1806640625)(8,0.2626953125)(9,0.3427734375)(10,0.2939453125)(11,0.4033203125)(12,0.376953125)(13,0.4404296875)(14,0.5615234375)(15,0.6650390625)(16,0.8759765625)(17,0.9833984375)(18,0.998046875)(19,1.0)(20,1.0)(21,0.9990234375)(22,0.98828125)(23,0.978515625)(24,0.873046875)(25,0.9033203125)(26,0.9296875)(27,0.9794921875)(28,0.935546875)(29,0.9609375)(30,0.9482421875)(31,0.98828125)(32,0.994140625)
    };
\addplot[
    color=black,
    mark=diamond*,
    mark size=0.5,
    very thick,
    ]
    coordinates {
    (1,0.140625)(2,0.021484375)(3,0.033203125)(4,0.033203125)(5,0.0546875)(6,0.04296875)(7,0.037109375)(8,0.0625)(9,0.0888671875)(10,0.091796875)(11,0.1650390625)(12,0.1552734375)(13,0.22265625)(14,0.3642578125)(15,0.494140625)(16,0.8076171875)(17,0.9638671875)(18,0.998046875)(19,1.0)(20,1.0)(21,0.9990234375)(22,0.9833984375)(23,0.97265625)(24,0.853515625)(25,0.861328125)(26,0.8984375)(27,0.96875)(28,0.8837890625)(29,0.916015625)(30,0.9111328125)(31,0.9853515625)(32,0.9912109375)
    };
    \legend{FFN1: All languages, FFN1: Highest activation rate for one language, FFN2: All languages, FFN2: Highest activation rate for one language}
\end{axis}
\end{tikzpicture}
  }
  \caption{Number of activated neurons in different conformer blocks. It is possible to prune 20\% - 80\% of the neurons for the first 12 layers, depending on the location of the adapter.}
    \label{fig:activation}
\end{figure}

We use the best system, TPA(1024), and compute the activation statistics based on the training set which covers all 102 languages.
From Figure~\ref{fig:activation}, clearly, higher layer requires a large adapter and neurons are mostly activated.
For the lower layer, the conclusion is mixed.
For the first FFN block, about 10\% to 20\% of the neurons could be pruned.
For the second FFN block, about 60\% to 80\% of the neurons could be pruned.
Our observation matches studies from previous work which claims adapters are important for higher layers~\cite{houlsby2019parameter}.
Indeed, when the task is changed from pre-training (MLM) to finetuning (ASR), the upper layer needs more adaptation.
Also, there are not much differences on the activation statistics for all languages and highest activation rate for single language.

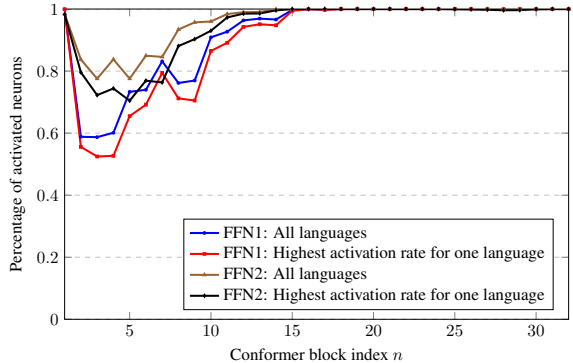
\begin{figure}[t]
  \centering
  \resizebox{0.9\columnwidth}{!}{
  \pgfplotsset{compat=1.3}
\begin{tikzpicture}
\begin{axis}[
    xlabel={Conformer block index $n$},
    ylabel={Percentage of activated neurons},
    ymin=0, ymax=1,
    xmin=1, xmax=32,
    legend pos=south east,
    ymajorgrids=true,
    grid style=dashed,
    legend cell align=left,
    width=12cm,height=8cm
]
\addplot[
    color=blue,
    mark=*,
    mark size=0.5,
    very thick,
    ]
    coordinates {
(1,0.9991861979166666)(2,0.5882161458333334)(3,0.5870768229166666)(4,0.6012369791666666)(5,0.7330729166666666)(6,0.73974609375)(7,0.8313802083333334)(8,0.76171875)(9,0.76953125)(10,0.9088541666666666)(11,0.9269205729166666)(12,0.96337890625)(13,0.9690755208333334)(14,0.9661458333333334)(15,0.9972330729166666)(16,0.99951171875)(17,0.9982096354166666)(18,1.0)(19,1.0)(20,1.0)(21,1.0)(22,1.0)(23,1.0)(24,1.0)(25,1.0)(26,1.0)(27,0.99951171875)(28,0.9998372395833334)(29,0.9998372395833334)(30,0.9998372395833334)(31,1.0)(32,1.0)
    };
\addplot[
    color=red,
    mark=square*,
    mark size=0.5,
    very thick,
    ]
    coordinates {
(1,0.9991861979166666)(2,0.5556640625)(3,0.5247395833333334)(4,0.52685546875)(5,0.6549479166666666)(6,0.6917317708333334)(7,0.7941080729166666)(8,0.7120768229166666)(9,0.705078125)(10,0.8649088541666666)(11,0.8912760416666666)(12,0.9423828125)(13,0.9510091145833334)(14,0.9475911458333334)(15,0.9943033854166666)(16,0.99951171875)(17,0.9972330729166666)(18,1.0)(19,1.0)(20,1.0)(21,1.0)(22,1.0)(23,1.0)(24,1.0)(25,1.0)(26,0.99951171875)(27,0.99951171875)(28,0.99951171875)(29,0.9998372395833334)(30,0.9996744791666666)(31,0.9998372395833334)(32,1.0)
    };
\addplot[
    color=brown!80!black,
    mark=triangle*,
    mark size=0.5,
    very thick,
    ]
    coordinates {
(1,0.984375)(2,0.8372395833333334)(3,0.7755533854166666)(4,0.8377278645833334)(5,0.7755533854166666)(6,0.849609375)(7,0.845703125)(8,0.93408203125)(9,0.95703125)(10,0.9599609375)(11,0.98388671875)(12,0.98974609375)(13,0.9905598958333334)(14,0.9993489583333334)(15,0.9998372395833334)(16,1.0)(17,1.0)(18,1.0)(19,1.0)(20,1.0)(21,1.0)(22,1.0)(23,1.0)(24,1.0)(25,0.9996744791666666)(26,0.9998372395833334)(27,0.9996744791666666)(28,0.9986979166666666)(29,0.99951171875)(30,0.9996744791666666)(31,1.0)(32,1.0)
    };
\addplot[
    color=black,
    mark=diamond*,
    mark size=0.5,
    very thick,
    ]
    coordinates {
(1,0.9825846354166666)(2,0.7960611979166666)(3,0.7224934895833334)(4,0.7439778645833334)(5,0.70458984375)(6,0.7692057291666666)(7,0.7633463541666666)(8,0.8806966145833334)(9,0.9029947916666666)(10,0.9300130208333334)(11,0.9728190104166666)(12,0.9851888020833334)(13,0.9856770833333334)(14,0.99560546875)(15,0.9998372395833334)(16,1.0)(17,1.0)(18,1.0)(19,1.0)(20,1.0)(21,1.0)(22,1.0)(23,1.0)(24,1.0)(25,0.99951171875)(26,0.9988606770833334)(27,0.9978841145833334)(28,0.9959309895833334)(29,0.9964192708333334)(30,0.99951171875)(31,1.0)(32,1.0)
    };
    \legend{FFN1: All languages, FFN1: Highest activation rate for one language, FFN2: All languages, FFN2: Highest activation rate for one language}
\end{axis}
\end{tikzpicture}
  }
  \caption{Number of activated neurons within each FFN in different conformer blocks. Not much neurons could be pruned from the frozen encoder, except for layer 2-10.}
    \label{fig:activation2}
\end{figure}

Based on this observation, we want to know if the frozen encoder could be pruned since it also speeds up the inference process.
If we could shrink the size of FFNs for the upper layers, we could still reduce the computation costs. The result is demonstrated in Figure~\ref{fig:activation2}.
It is possible to prune 20\% to 40\% of the neurons from layer 2-10 without performance loss.

\begin{table}[hbt!]
\caption{Speech recognition results on SpeechStew and Fleurs. Results are reported in Word Error Rate (WER).}
\label{tab:asr}
\footnotesize
\centering
\begin{tabular}{@{}cccccc@{}}
\toprule
\multicolumn{2}{c}{Dataset}                                                                                 & Finetune & Serial(256) &  TPA(128) & TPA(256) \\ \midrule \midrule
\multirow{2}{*}{AMI}                                                           & \multicolumn{1}{c|}{IHM}   & {\bf 8.9} & 9.9    & 9.7  & 9.7      \\ \cmidrule(l){2-6} 
                                                                              & \multicolumn{1}{c|}{SDM1}  & {\bf 20.1} & 22.3   & 21.2  & 20.9     \\ \midrule
\multicolumn{2}{c|}{\begin{tabular}[c]{@{}c@{}}Common\\ Voice\end{tabular}}                                 & {\bf 9.0} & 9.8    & 9.4 & {\bf 9.0}      \\ \midrule
\multirow{2}{*}{LibriSpeech}                                                   & \multicolumn{1}{c|}{clean} & 2.0 &2.0         & {\bf 1.9} & {\bf 1.9} \\ \cmidrule(l){2-6} 
                                                                              & \multicolumn{1}{c|}{other} & 4.0    &3.8      & 3.8 & {\bf 3.7} \\ \midrule
\multirow{2}{*}{\begin{tabular}[c]{@{}c@{}}Switchboard\\ /Fisher\end{tabular}} & \multicolumn{1}{c|}{SWBD}  & {\bf 4.5} & 4.6  & 5.6 & {\bf 4.5} \\ \cmidrule(l){2-6} 
                                                                              & \multicolumn{1}{c|}{CH}    & {\bf 8.1} & 9.2   & 8.9 & 8.9      \\ \midrule
\multicolumn{2}{c|}{WSJ}                                                                                    & 1.5 &1.6        & 1.6 & {\bf 1.4}      \\ \midrule
\multicolumn{2}{c|}{Tedlium}                                                                                & {\bf 5.7} & 6.3          & 6.4 & 6.1      \\ \midrule \midrule
\multicolumn{2}{c|}{Fleurs}                                                                                 & {\bf 18.0} &  19.4       & 19.1 & 18.7     \\ \bottomrule
\end{tabular}
\end{table}

\subsection{Studies on different tasks and datasets}

Next, we test the proposed TPA approach on different speech tasks and datasets to demonstrate whether this could become a general approach: serving the same frozen pre-trained encoder with different task-dependent adapter.
Our studies focus on two challenging scenarios: Speech Recognition and Speech Translation.
For Speech Recognition, we test on one additional public benchmark SpeechStew, in addition to the multilingual dataset Fleurs.
SpeechStew~\cite{speechstew} is a combination of seven public corpora, including AMI, Common Voice, English Broadcast News,
 LibriSpeech, Switchboard/Fisher,
TED-LIUM v3 and Wall Street Journal.
All of them are English datasets with different recording environments.
The speech recognition results are reported on Table~\ref{tab:asr}.
TPA achieves same or better results as full network finetuning on half of the datasets and slightly worse on challenge ones while only 2\% of parameters are finetuned.

We also test our approach on Speech Translation tasks and CoVoST 2 \cite{wang2020covost} is used to benchmark multilingual speech translation. 
Following \cite{bapna2022mslam}, we choose multilingual XX-to-English task which covers 21 source languages. 
The amount of training data varies between 1 - 264 hours depending on the source language.
Results are reported in Table~\ref{tab:covost2}.
We only observe ~0.9 BLEU score difference between full finetuning and the proposed TPA approach.

\begin{table}[htb!]
\caption{Performance on Covost 2. Results are reported by BLEU score.}
\label{tab:covost2}
\centering
\begin{tabular}{c|c|c}
 \toprule
 System & \% of updated paramters  & BLEU \\
 \midrule
 Finetune All & 100.0\% & \textbf{28.7} \\
 Serial(512) & 1.4\% & 27.3 \\
 \midrule
 TPA(256) & \textbf{1.4\%} & 27.8 \\
 \bottomrule
\end{tabular}
\end{table}

\section{Summary}
\label{sec:sum}

We propose a new variation of the residual adapter, Two Parallel Adapters (TPA), for the Conformer architecture, which is widely used in many speech applications.
TPA adapts both feed-forward networks of a conformer with parallel residual adapters.
Given a pre-trained giant encoder, we attach TPA and a randomly initialized decoder.
During adaptation, only the TPA and the decoder are updated.
Compared to full network finetuning, comparative studies reveal that on multilingual speech recognition tasks, TPA matches the performance of full-network finetuning, but only updates 11\% of the parameters. Furthermore, TPA outperforms both serial and parallel adapters proposed in the literature.
Ablation studies show TPA could also be pruned, especially for lower layers, without performance loss.
Finally, extensive studies suggest that TPA is a general approach applicable to a lot of downstream speech tasks, including speech recognition and translation.

In the future, we plan to focus more on improving the performance of small TPAs.
One potential direction is to perform iterative pruning, which prunes non-activated neurons iteratively.
We could even introduce some sparsity loss on the adapter neurons to make ReLU sparse so neurous could be pruned.
Another possibility is to perform knowledge distillation from a large TPA to a small TPA.

\bibliographystyle{IEEEbib}
\bibliography{refs}

\end{document}